\documentclass[12pt]{article}
\pdfoutput=1

\usepackage{amssymb}
\usepackage{citesort}
\usepackage{amsmath}
\usepackage{url}
\usepackage{graphicx}
\usepackage{subfigure}

\def\pa{\partial}

\def\hb{\hbar}

\def\ii{\textrm i}

\def\ve{\varepsilon}

\newcommand{\db}{de$\,$Broglie}

\newcommand{\dbb}{de$\,$Broglie-Bohm}
\newcommand{\Dbb}{De$\,$Broglie-Bohm}

\begin{document}
\vspace*{1.0cm}
\noindent
{\bf
{\large
\begin{center}
Quantum non-equilibrium and relaxation to equilibrium for a class of \dbb-type theories
\end{center}
}
}
\vspace*{.5cm}
\begin{center}
Samuel Colin{\footnote{PIAF Postdoctoral Research Fellow.}}
\end{center}

\begin{center}
Centre for Quantum Dynamics, Griffith University,\\
Brisbane Queensland 4111, Australia.\\
E--mail: \texttt{s.colin@griffith.edu.au}.
\end{center}

\begin{center}
Ward Struyve{\footnote{Postdoctoral Fellow FWO.}}
\end{center}

\begin{center}
Institute of Theoretical Physics, K.U.Leuven,\\
Celestijnenlaan 200D, B--3001 Leuven, Belgium.{\footnote{Corresponding address.}}\\
Institute of Philosophy, K.U.Leuven,\\
Kardinaal Mercierplein 2, B--3000 Leuven, Belgium.\\
E--mail: \texttt{Ward.Struyve@fys.kuleuven.be}.
\end{center}

\begin{abstract}
\noindent
The \dbb\ theory is about non-relativistic point-particles that move deterministically along trajectories. The theory reproduces the predictions of standard quantum theory, given that the distribution of particles over an ensemble of systems, all described by the same wavefunction $\psi$, equals the quantum equilibrium distribution $|\psi|^2$. Numerical simulations by Valentini and Westman have illustrated that non-equilibrium particle distributions may relax to quantum equilibrium after some time. Here we consider non-equilibrium distributions and their relaxation properties for a particular class of trajectory theories, first studied in detail by Deotto and Ghirardi, that are empirically equivalent to the \dbb\ theory in quantum equilibrium. For the examples of such theories that we consider, we find a speed-up of the relaxation compared to the ordinary \dbb\ theory. Hence non-equilibrium predictions that depend strongly on relaxation properties, such as those studied recently by Valentini, may vary for different trajectory theories. As such these theories might be experimentally distinguishable. 
\end{abstract}

\renewcommand{\baselinestretch}{1.1}
\bibliographystyle{unsrt}

\section{Introduction}
In the \dbb\ theory \cite{bohm93,holland93b,durr09} an individual closed system is described by its wavefunction that satisfies the Schr\"odinger equation and by particle positions that move deterministically along trajectories with a velocity that depends on the wavefunction. The theory reproduces the predictions of standard quantum theory, given that the distribution of particle positions over an ensemble of systems, all described by the same wavefunction $\psi(x,t)$, is given by $|\psi(x,t)|^2$. The dynamics is such that if the distribution of particle positions equals $|\psi(x,t_0)|^2$ at a certain time $t_0$, then it equals $|\psi(x,t)|^2$ at other times $t$, a property called {\em equivariance} \cite{durr92}. The distribution $|\psi|^2$ plays the role of an equilibrium distribution (see for example \cite{bohm53b,valentini91a,durr92}), similar to that of thermal equilibrium in classical statistical mechanics, and is called the {\em quantum equilibrium distribution}.

Current day experiments yield a very good confirmation of standard quantum theory and hence of quantum equilibrium. This could be accounted for by a Boltzmann-type argument: as shown by D\"urr {\em et al.}\ \cite{durr92} most initial configurations of the universe (relative to the natural measure $|\Psi(X)|^2dX$, with $\Psi$ the wavefunction of the universe) yield the quantum equilibrium distribution $|\psi(x)|^2$ for actual ensembles described by the effective wavefunction $\psi(x)$. In addition, it is to be expected that non-equilibrium distributions will tend to equilibrium and stay close to it for a reasonable time in ``most'' cases.{\footnote{As explained in detail in Section \ref{relaxation}, some distributions do not evolve to quantum equilibrium. For some wavefunctions there even is never relaxation to equilibrium. So the ``most'' should in the first place refer to some natural measure $\mu(d\psi)$ on Hilbert space. For example, considering product states $\psi(x_1)\dots \psi(x_N)$ (with $N$ large), it probably holds that most non-equilibrium distributions (understood as empirical distributions) will tend to evolve to equilibrium, where ``most'' refers to the measure $|\psi(x_1)\dots \psi(x_N)|^2dx_1 \dots dx_N \mu(d\psi)$. In the following we will not attempt to make these statements more precise.}} This was illustrated by Valentini and Westman \cite{valentini05}, who presented a numerical simulation of such a relaxation to equilibrium for a free particle in a two-dimensional box. For a particular initial non-equilibrium distribution they found a rapid relaxation to quantum equilibrium (with relaxation time-scale roughly given by $\hbar^2/\varepsilon m^{1/2} (\Delta E)^{3/2}$, where $\varepsilon$ is the coarse-graining length that is used in comparing the coarse-grained non-equilibrium and equilibrium distributions, $m$ the mass of the particle and $\Delta E$ the energy spread of the wavefunction).

Non-equilibrium distributions have been studied in detail by Valentini, who suggested possible domains where such non-equilibrium distributions might occur and how they might be detected, see for example \cite{valentini91a,valentini92,valentini04a,valentini07}. In particular, Valentini suggested to look for non-equilibrium in astrophysical and cosmological scenarios. Assuming that the universe started in a special state of non-equilibrium (just like our universe seems to have started from a very special state of non-equilibrium on the classical statistical level), he considered circumstances under which this non-equilibrium might be preserved and even be transferred to macroscopic scales. A priori, any non-equilibrium seems possible. Nevertheless, in the case of a scalar field on expanding space, Valentini was able to come up with a quantitative prediction \cite{valentini08a,valentini08b}. He showed that relaxation to equilibrium is expected to be suppressed for a specific range of modes. One of the possible consequences is a correction to the predictions for the temperature fluctuations of the cosmic microwave background in the context of inflation theory.

In this paper we consider a certain class of trajectory theories that, just as the \dbb\ theory, leave $|\psi|^2$ equivariant. In this class of theories, that was studied in detail by Deotto and Ghirardi \cite{deotto98}, the velocity field of the particles is changed by an additive term compared to that of the \dbb\ theory. From the point of view of these theories, standard quantum theory emerges as an effective theory describing the state of equilibrium. As such they may only be distinguished in quantum non-equilibrium. The goal of this paper is to examine quantum non-equilibrium and possible relaxation to equilibrium for these theories, by means of numerical simulations. 

In the examples of such theories that we study, the relaxation proceeds more rapidly than in the ordinary \dbb\ theory. This was as expected, since the additional term in the velocity field in general just adds to the irregularity of the motion and hence to its mixing-like behaviour. In particular, in the case of the ordinary \dbb\ theory, it was reported before that nodes in general induce chaotic behaviour for trajectories that come in their neighbourhood (see for example \cite{frisk97,wu99,valentini05,wisniacki05,wisniacki06,efthymiopoulos07,efthymiopoulos09}).{\footnote{See for example \cite{efthymiopoulos09} or \cite{wyatt05} for a collection of references on chaos in the \dbb\ theory.} In the alternative theories under consideration, the additional term in the velocity fields tends to lead to an overall increase of its amplitude, causing the particles to travel larger distances, compared to the ordinary \dbb\ theory, in a given time interval. As such, particles tend to reach nodes sooner and undergo the chaotic motion sooner. 

Non-equilibrium predictions that depend strongly on relaxation properties, such as those studied recently by Valentini, may hence vary for different trajectory theories. As such these theories might be experimentally distinguishable.

The outline of the paper is as follows. In Section \ref{dbb}, we start with reviewing this class of \dbb-type theories. In Section \ref{nonequilibrium}, we consider non-equilibrium distributions and some issues that are relevant to possible relaxation. In Section \ref{simulations}, we present numerical simulations of relaxation for a number of \dbb-type theories, for a particle in a two-dimensional box (just as in the simulations of Valentini and Westman), finding a correlation between relaxation time and the choice of dynamics.

\section{\Dbb\ theory}\label{dbb}
\subsection{Standard velocity field}
In the \dbb\ theory, an individual closed system is described by its wavefunction $\psi(x,t)$, with $x=({\bf x}_1,\dots,{\bf x}_N) \in {\mathbb R}^{3N}$, which satisfies the non-relativistic Schr\"odinger equation
\begin{equation}
\ii \hb \frac{\pa \psi(x,t)}{\pa t} = - \sum^N_{k=1} \frac{\hb^2}{2m_k}\nabla^2_{k}\psi(x,t) + V(x) \psi(x,t) \,,
\label{1}
\end{equation}
and by particle positions ${\bf X}_1(t),\dots,{\bf X}_N(t)$, whose configuration $X(t)=({\bf X}_1(t),\dots,{\bf X}_N(t))$ satisfies the {\em guidance equation}
\begin{equation}
\frac{d X(t) }{dt} = v_s(X(t),t)\,,
\label{2}
\end{equation}
where the velocity field $v_s$ is given by
\begin{equation}
v_s(x,t) = \frac{j_s(x,t)}{|\psi(x,t)|^2}  \,,
\label{3}
\end{equation}
with $j_s = ({\bf j}_{s,1},\dots,{\bf j}_{s,N})$ the standard quantum probability current, given by 
\begin{equation}
{\bf j}_{s,k} = \frac{\hbar}{m_k} {\textrm{Im}}\left(\psi^* {\boldsymbol \nabla}_k  \psi \right)=\frac{1}{m_k}   {\boldsymbol \nabla}_k S |\psi|^2 \,, \qquad \psi = |\psi| \exp(\ii S/\hb)\,, 
\label{3.1}
\end{equation}
and which satisfies the continuity equation
\begin{equation}
\frac{\pa |\psi|^2}{\pa t} +  \nabla \cdot j_s = 0 
\label{4}
\end{equation}
as a consequence of the Schr\"o\-din\-ger equation.

An arbitrary distribution $\rho(x,t)$ transported along the \db-Bohm trajectories satisfies the continuity equation
\begin{equation}
\frac{\pa \rho}{\pa t} +  \nabla \cdot (v_s \rho)  = 0 \,.
\label{5}
\end{equation}
Also $|\psi|^2$ satisfies this continuity equation, because of \eqref{4}. This implies equivariance: if the distribution $\rho$ equals $|\psi|^2$ at a certain time, then they are equal at all time (and $|\psi|^2$ is actually the only equivariant distribution that is also a suitably local functional of $\psi$ \cite{goldstein07}). It is the generalization of the property of stationarity in classical statistical mechanics.

As mentioned in the introduction $|\psi|^2$ plays the role of an equilibrium distribution, called the {\em quantum equilibrium distribution}. Given the quantum equilibrium distribution and the fact that measurement results are generally ultimately recorded in positions of macroscopic pointers, like instrument needles, computer outprint, etc., it almost follows immediately that the \dbb\ theory reproduces the standard quantum mechanical predictions (see for example \cite{bohm52b,bohm93,holland93b,durr033,durr09}).

\subsection{Alternative velocity fields}
The flow defined by \eqref{2} and \eqref{3} is not the only one for which the distribution $|\psi|^2$ is equivariant. The continuity equation for $|\psi|^2$ still holds if the current $j_s = v_s |\psi|^2$ is replaced by 
\begin{equation}
j = j_s + j_a \,,
\label{8}
\end{equation}
where $\nabla \cdot j_a=0$. As a result, for the velocity field
\begin{equation}
v = \frac{ j}{|\psi|^2} = v_s  + \frac{j_a}{|\psi|^2}  \,,
\label{9}
\end{equation}
the distribution $|\psi|^2$ will also be equivariant. As such one can consider alternative theories, that just as the ordinary \dbb\ theory, agree with standard quantum theory in quantum equilibrium. 

This ambiguity in the choice of possible guidance equations was first studied in detail by Deotto and Ghirardi \cite{deotto98}, who found that the requirement of Galilean covariance is insufficient to determine the guidance equation uniquely. Uniqueness could be obtained by extra requirements \cite{durr92,durr09} or by different ones \cite{holland99,holland03b,struyve03b,wiseman07}. These uniqueness results merely indicate the possible naturalness of certain guidance equations.

An interesting example of this ambiguity turns up in the context of the Pauli equation, which describes non-relativistic spin-1/2 particles \cite{bohm93}. Considering a single particle, one possible current is given by 
\begin{equation}
{\bf j}_s = \frac{\hbar}{m} {\textrm{Im}}\left( \Psi^\dagger {\boldsymbol \nabla} \Psi  \right) \,.
\label{10}
\end{equation}
When the magnetic field is negligible and $\Psi$ is a spin eigenstate, that is, $\Psi({\bf x},t) = \psi({\bf x},t) \chi$, where $\chi$ is a constant spinor, the Pauli equation will imply the non-relativistic Schr\"odinger equation for $\psi$ and the current will reduce to the one in \eqref{3.1}. Another natural current is obtained by adding the divergence-free term
\begin{equation}
{\bf j}_a = \frac{\hbar}{2m} {\boldsymbol \nabla}  \times (\Psi^\dagger  {\boldsymbol \sigma} \Psi) 
\label{11}
\end{equation}
to ${\bf j}_s$. For a spin eigenstate this term survives, so that the latter current yields the velocity field
\begin{equation}
{\bf v} = {\bf v}_s  + \frac{\hbar}{2m|\psi|^2}   {\boldsymbol \nabla}  \times \left(|\psi|^2 \chi^{\dagger} {\boldsymbol \sigma}  \chi  \right) = \frac{1}{m} {\boldsymbol \nabla} S +\frac{\hbar}{2m|\psi|^2}   {\boldsymbol \nabla}  \times \left(|\psi|^2 \chi^{\dagger} {\boldsymbol \sigma}  \chi \right)\,.
\label{12}
\end{equation}
This velocity field is very natural when one considers the non-relativistic spin-1/2 theory as the limiting case of the relativistic Dirac theory \cite{bohm93}. The natural current and velocity field for the Dirac theory namely yield the spin-term in the non-relativistic limit. The trajectories for this velocity field have been studied in for example \cite{holland03a,colijn02,colijn03a,colijn03b}. (For other spins a similar ambiguity can be considered \cite{struyve05}.)   

Note that one can have other trajectory theories in which the velocity field is not of the form \eqref{10}, but which are still empirically equivalent with the \dbb\ theory in quantum equilibrium, see for example \cite{holland98,goldstein042}. There even exist such theories where the equilibrium distribution is in general different from $|\psi|^2$ \cite{struyve09b}.

\subsection{Some properties of the alternative velocity fields}
The velocity fields $v$ given in \eqref{9} are not defined at the nodes of the wavefunction (that is, points where $\psi=0$). In general they also diverge near nodes. Nevertheless, in the case of the ordinary \dbb\ velocity field, the set of initial configurations that run into a node has $|\psi|^2$-measure zero \cite{berndl93,teufel04}. Probably similar results hold for a reasonable class of alternative velocity fields.

In the case of the ordinary \dbb\ velocity field, the nodes are the only potential source of vorticity, while the alternative velocity fields yield vorticity even away from nodes. This can be seen as follows. Consider a single particle for simplicity. The average vorticity $\omega_\Sigma$ over a surface $\Sigma$ with boundary $\partial \Sigma$ is given by 
\begin{equation}
\omega_\Sigma = \frac{1}{ A(\Sigma)} \oint_{\partial \Sigma} {\bf v} \cdot d{\bf l} \,,
\label{20}
\end{equation}
where $\oint_{\partial \Sigma} {\bf v} \cdot d{\bf l}$ is the circulation of the velocity field and $ A(\Sigma)$ is the area of the surface. If the surface does not contain any nodes, Stokes' theorem can be applied to write $\oint_{\partial \Sigma} {\bf v} \cdot d{\bf l} = \int_\Sigma {\boldsymbol \omega} \cdot d{\boldsymbol \Sigma}$, where ${\boldsymbol \omega} = {\boldsymbol \nabla} \times {\bf v}$ is the vorticity vector. This vorticity vector is not defined at nodes (though alternatively one could define it in a distributional sense, see for example \cite{takabayasi83} for the case of the ordinary \dbb\ theory).

In the case of the ordinary \dbb\ theory the velocity field is given by ${\bf v}_s = {\boldsymbol \nabla} S/m$, so that the corresponding vorticity vector is zero. Hence if the surface $\Sigma$ does not contain any nodes, then the average vorticity over that surface is zero. The average vorticity can only be non-zero when the surface contains nodes. In that case the vorticity is quantized \cite{dirac31,hirschfelder74}. This is because the circulation is given by $\oint_{\partial \Sigma} {\bf v} \cdot d{\bf l} = \oint_{\partial \Sigma} {\boldsymbol \nabla} S  \cdot d{\bf l}/m = nh/m$, with $n$ integer, since $S$ is multivalued, where different values of $S$ at a point can only differ by an integer multiple of $2\pi \hbar = h$ (since $\psi$ is single-valued). 

Note that in the case the particle is moving in an external magnetic field ${\bf B} = {\boldsymbol \nabla} \times {\bf A}$, the standard velocity field becomes ${\bf v}_s = {\boldsymbol \nabla} S/m - e{\bf A}/mc$, so that the vorticity vector obtains a contribution $-e{\bf B}/mc$.

In the case of alternative velocity fields, the vorticity vector ${\boldsymbol \omega}$ is in general different from zero, so that there is vorticity even away from nodes. 

In the ordinary \dbb\ theory, the chaotic motion near nodes probably finds its origin in the vorticity, together with the divergent behaviour of the velocity field. In the case of alternative velocity fields similar features hold even in the neighbourhood of quasi-nodes (where $|\psi|$ becomes very small). In this context, it would therefore be interesting to study the implications of vorticity, possibly combined with nodal or quasi-nodal structure, for possible chaotic behaviour.

\section{Quantum non-equilibrium and relaxation to quantum equilibrium}\label{nonequilibrium}
\subsection{Coarse-graining}\label{coarsegraining}
For an arbitrary distribution $\rho$ one can introduce the quantity $f=\rho/|\psi|^2$. Because $\rho$ and $|\psi|^2$ satisfy the same continuity equation (with velocity field $v$), one has that $\partial f / \partial t + v \cdot \nabla f = 0$. This means that $f$ is conserved along trajectories, that is, $f(X(t),t) = f(X(0),0)$ for all time $t$ \cite{bohm53b,valentini91a,valentini92,valentini05}. As such it would seem that $\rho$ can never really relax to $|\psi|^2$ because their ratio is conserved along trajectories. However, relaxation should be understood in the sense that $\rho$ goes to $|\psi|^2$ on a coarse-grained level \cite{valentini91a,valentini92,valentini05}. That is, if we consider the coarse-grained densities ${\bar \rho}$ and ${\overline{|\psi|^2}}$, whose value is obtained by averaging over non-overlapping cells (so that their value is constant over those cells), then ${\bar \rho}/{\overline{|\psi|^2}}$ is not necessarily conserved along the trajectories, so that ${\bar \rho}$ may relax to ${\overline{|\psi|^2}}$. The coarse-graining could be understood as corresponding to a finite accuracy of physical measurements. 

The deviation of $\rho$ from $|\psi|^2$ can be quantified by means of minus the relative entropy of $\rho$ with respect to $|\psi|^2$ \cite{valentini91a,valentini92,valentini05}:
\begin{equation}
H = \int dx \rho \ln \left( \rho / |\psi|^2 \right) \,.
\label{30}
\end{equation}
Valentini called this quantity $H$ in analogy with the $H$-function in classical statistical mechanics. It is strictly positive and zero for the quantum equilibrium distribution. However, since this quantity is conserved in time, just like $f$, it is not really suitable to quantify relaxation. Instead the coarse-grained $H$-function 
\begin{equation}
{\bar H} = \int dx {\bar \rho} \ln \left({\bar \rho} / {\overline{|\psi|^2}} \right) 
\label{31}
\end{equation}
should be considered.

\subsection{Relaxation}\label{relaxation}
While we typically expect a non-equilibrium distribution to relax to equilibrium (on a coarse-grained level), this will definitely not happen always. That is, while we typically expect the coarse-grained $H$-function, ${\bar H}$, to decrease over time, it may also increase over some time, or stay constant. This depends on both the initial distribution and the initial wavefunction. Let us expand on this.

First of all, for a given wavefunction, not every initial distribution $\rho(x)$ will have relaxed to the quantum equilibrium distribution $|\psi(x,T)|^2$ in a given time $T$. Just consider some distribution $\rho(x) \neq |\psi(x,T)|^2$ and evolve that distribution backwards in time along the \dbb\ trajectories. The resulting distribution $\rho(x,-T)$ will then lead to the non-equilibrium distribution $\rho(x)$ after time $T$. The distribution might of course relax at a later time.

Second, for some wavefunctions there will even be no relaxation, regardless of the initial distribution. For example, in the case of the ordinary \dbb\ theory there will not be relaxation for a plane wave or a Gaussian wavefunction (see \cite{colijn04} for further examples). For a single-particle plane wave any distribution will just be translated over time. For a Gaussian wavefunction any distribution will either spread or contract over time, depending on whether the Gaussian itself is spreading and contracting. Note that relaxation might still occur for alternative velocity fields.

Third, if there is relaxation of a distribution $\rho$ for a particular wavefunction, then by time reversal invariance there exists a wavefunction for which the time evolution of the distribution becomes time reversed \cite{valentini05}. The \dbb\ dynamics has the time-reversal symmetry $(X(t),\psi(x,t)) \to (X(-t),\psi^*(x,-t))$, which reverses the direction in which the trajectories are traversed. As a result, an initial distribution $\rho(x)$ whose time evolution is given by $\rho(x,t)$ under the dynamics determined by $\psi(x,t)$, will evolve according to $\rho(x,-t)$ under the dynamics determined by $\psi^*(x,-t)$.

Lastly, if the wavefunction is periodic and if the velocity field inherits this periodicity (which is the case for the theories for which we perform the numerical simulations), the trajectories typically recur. This follows by applying the standard Poincar\'e recurrence theorem. One formulation of the theorem states that for a probability space $(\Gamma,{\mathcal B},\mu)$ and $f$ a measure preserving map, that is, $\mu \circ f^{-1} = \mu$, one has that for $A \in {\mathcal B}$, for almost every $x \in A$ (with respect to the measure $\mu$), $f^n(x) \in A$ for infinitely many $n \in \mathbb{N}$ \cite[p.\ 26]{walters82}. One can apply this theorem for $\Gamma \subset {\mathbb R}^{3N}$ the configuration space, $\mu(dx)$ the measure $|\psi(x,0)|^2 dx$ and $f$ the map $x_{\tau}$, where $x_t$ is the flow map associated to the velocity field \eqref{9} and $\tau$ the period of the wavefunction. Equivariance means that $\mu \circ x^{-1}_t(dx) = |\psi(x,t)|^2 dx$. Hence $\mu \circ x^{-1}_{\tau}(dx) = |\psi(x,\tau)|^2 dx = |\psi(x,0)|^2 dx = \mu (dx)$, so that the map $x_{\tau}$ is measure preserving. Note further that $x_{t+\tau}=x_t \circ x_\tau$. This can be seen as follows. First, $Y(t) \equiv x_{t+\tau}(X)$, $X \in \Gamma$, is a possible trajectory, since
\begin{equation}
\frac{dY(t)}{dt} =\frac{dx_{t + \tau}(X)}{d(t+\tau)}= v(x_{t + \tau}(X),t + \tau)=v(Y(t),t)\,,
\label{32}
\end{equation} 
with initial configuration $x_{\tau}(X) = X(\tau)$. Since the trajectory $x_t \circ x_\tau (X)$ has the same initial configuration $x_\tau(X)$, if follows that $x_{t + \tau} = x_t \circ x_\tau$.{\footnote{Note that, except in one dimension \cite{goldstein99}, the trajectories in general do not inherit the periodicity of the wavefunction, that is, in general $x_{t + n\tau} \neq x_t$. Note also that the maps $x_t$ in general do not form a group.}} Hence $x^n_{\tau} = x_{n \tau}$. Application of the Poincar\'e recurrence theorem now yields that for $A \in {\mathcal B}$, for almost every $x \in A$, $x_{n \tau}(x) \in A$ for infinitely many $n \in \mathbb{N}$, that is, the trajectory recurs an infinite number of times with the recurrence times given by an integer multiple of the period of the wavefunction. 

This recurrence theorem has the following implication for the simulations. The time evolution of a non-equilibrium density is simulated by considering its evolution only along a finite number of trajectories. From the recurrence theorem it follows that such a collection of trajectories will typically recur. This means that if we could run the simulation for arbitrary times, there would be times at which the sampled non-equilibrium distribution would recur. 

While the recurrence theorem indicates that distributions may evolve away from quantum equilibrium, it is expected that for ``most'' distributions the recurrence time is much larger than the time spent near equilibrium.

\section{Numerical simulations}\label{simulations}
\subsection{Particle in a two dimensional square box}
We consider a particle that moves in two dimensions and that is confined to a two-dimensional square box of side $\pi$ by an infinite potential well ($\hbar$, the mass $m$ and the length unit are all put to 1). As such, the wavefunction is given by a superposition of the energy eigenfunctions
\begin{equation}
\phi_{mn}(x_1,x_2) =  \frac{2}{\pi} \sin(mx_1)\sin(nx_2) 
\label{50}
\end{equation}
with energy eigenvalues $E_{mn} = (m^2 + n^2)/2$, where $m$ and $n$ are strictly positive integers. 

We first consider a superposition of the four lowest energy eigenstates, with equal weights and different phases $\theta_{mn}$:{\footnote{For the record, the phases are given by $\theta_{11} =1.1525988926093297$, $\theta_{12} = 4.2775762116024665$, $\theta_{21} =2.1660329888555025$, $\theta_{22} =2.8960554218806349$. They are obtained from \cite{valentini05}, where they were randomly generated.}
\begin{equation}
\psi_1(x_1,x_2,t) =   \sum^2_{m,n=1} \frac{1}{2}e^{\ii \theta_{mn}}  \phi_{mn}(x_1,x_2) e^{-\ii E_{mn}t } = \sum^2_{m,n=1} \frac{1}{\pi} \sin(mx_1)\sin(nx_2) e^{\ii (\theta_{mn} - E_{mn}t) } \,.
\label{51}
\end{equation}
The wavefunction is periodic, with period $4\pi$. The corresponding density $|\psi_1(x_1,x_2,0)|^2$, at time $t=0$, is given in figure \ref{psisq}. The wavefunction has only one node (within the box), whose trajectory is given in figures \ref{trajectories1} and \ref{trajectories2}.

\begin{figure}
\begin{center}
\includegraphics[height=5cm]{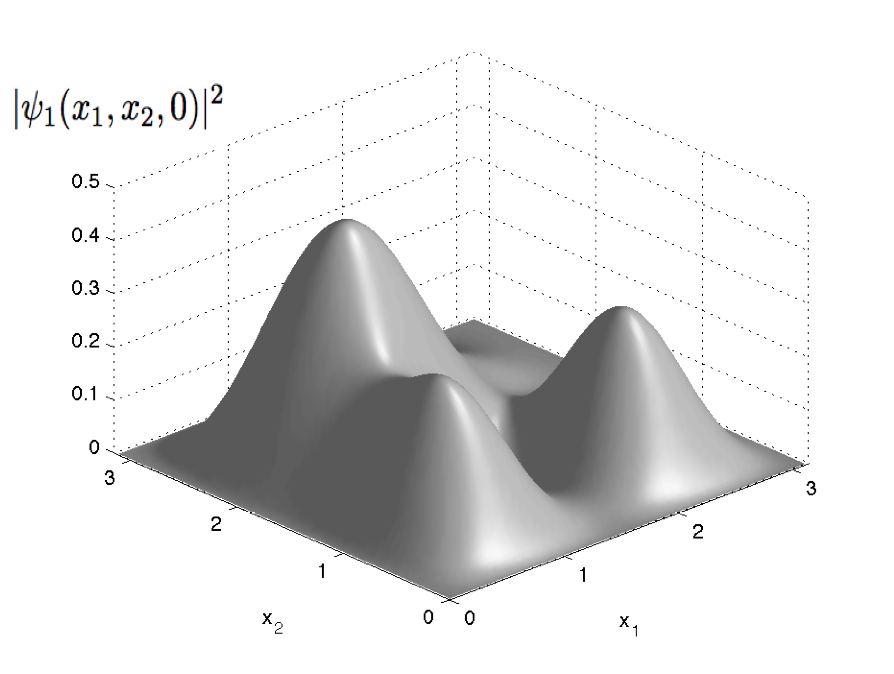}
\includegraphics[height=5cm]{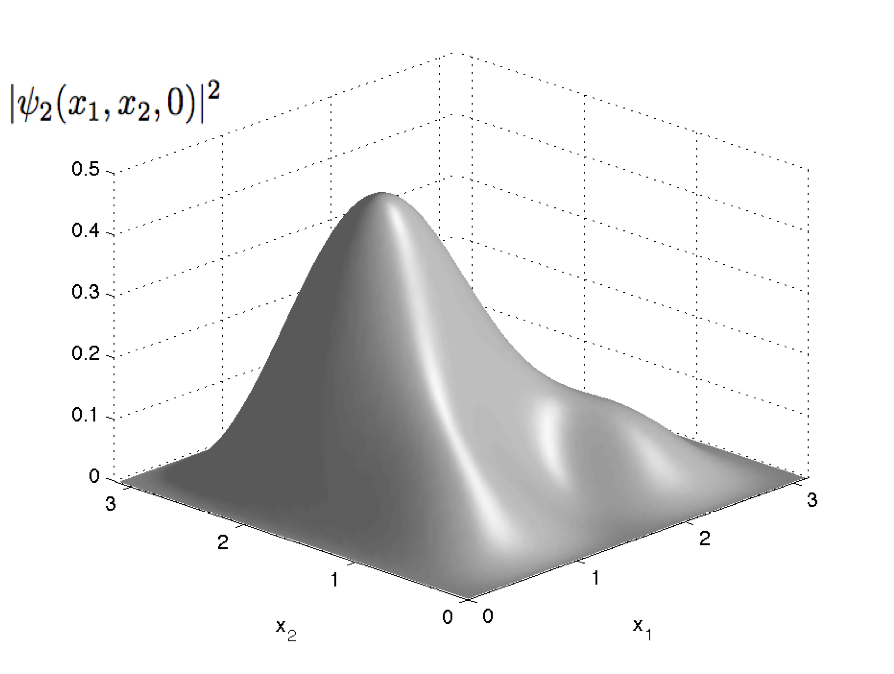}
\caption{The equilibrium densities at time $t=0$, for the wavefunctions $\psi_1$ and $\psi_2$, given respectively in equations \eqref{51} and \eqref{51.1}.}
\label{psisq}
\end{center}
\end{figure}

We also consider the following wavefunction 
\begin{equation}
\psi_2(x_1,x_2,t) = \frac{\sqrt{3}}{2} e^{\ii\theta_{11}} \phi_{11}(x_1,x_2) e^{ - \ii  E_{11}t } + \sum^2_{\begin{subarray}{l} m,n=1\\ m+n >2 \end{subarray}} \frac{1}{2\sqrt{3}} e^{\ii \theta_{mn}} \phi_{mn}(x_1,x_2) e^{- \ii E_{mn}t } \,,
\label{51.1}
\end{equation}
which also has period of $4\pi$, but which doesn't have any nodes. The corresponding density $|\psi_2(x_1,x_2,0)|^2$ is given in figure \ref{psisq}.

In two dimensions a divergence-free two-vector field $a_i(x_1,x_2)$ can always be written as $a_i = \ve_{ij} \pa_j f(x_1,x_2)$, where $\ve_{ij}$ is anti-symmetric with $\ve_{12}=1$ (as a consequence of Poincar\'e's lemma), so that the velocity field of the particle can be put into the form
\begin{equation}
v_i(x_1,x_2,t) = v_{s,i}(x_1,x_2,t) + \mu \frac{\ve_{ij}\pa_j f(x_1,x_2,t)}{|\psi(x_1,x_2,t)|^2}\,,
\label{52}
\end{equation}
with $\mu$ a constant. We will consider a number of different values for $\mu$ and the following choices for the function $f$:{\footnote{Note that the velocity field \eqref{12}, for the case of a spin-eigenstate $\chi=(1\ 0)^T$ and for a factorizable wavefunction $\psi(x_1,x_2,x_3,t)=\varphi(x_1,x_2,t)\eta(x_3,t)$, reduces to 
\begin{equation}
v_1 = \frac{1}{m} \partial_1 S_{\varphi} + \frac{\hbar}{2m |\varphi|^2} \pa_2 |\varphi|^2 \,, \quad v_2 = \frac{1}{m} \partial_2 S_{\varphi} - \frac{\hbar}{2m |\varphi|^2} \pa_1 |\varphi|^2 \,, \quad   v_3 = \frac{1}{m} \partial_3 S_{\eta} \,, 
\label{52.2}
\end{equation}
where $S_{\varphi}$ and $S_{\eta}$ are respectively the phases of $\varphi$ and $\eta$. Hence in this case the motion in the $(x_1,x_2)$-plane decouples from that in the $x_3$ direction and the velocity field in the $(x_1,x_2)$-plane is of the form \eqref{52}, with $\mu = \hbar/2m$ and $f=f_1=|\varphi|^2$.}}
\begin{equation}
f_1 = |\psi|^2\,,\quad f_2 = \ve_{ij} \pa_i j_{s,j}\,,\quad f_3 = \pa_i j_{s,i} \,. 
\label{52.1}
\end{equation}
The value $\mu=0$ corresponds to the ordinary \dbb\ theory. The choices are such that there is no outgoing or incoming quantum probability flux at the boundary of the box (which means that the components of the currents normal to the boundary vanish). This means that there is no flow of $|\psi|^2$ in or out of the box. In figures \ref{trajectories1} and \ref{trajectories2} some trajectories are given for the wavefunction $\psi_1$ given in \eqref{51}. 

\begin{figure}
\centering
\includegraphics[width=\textwidth]{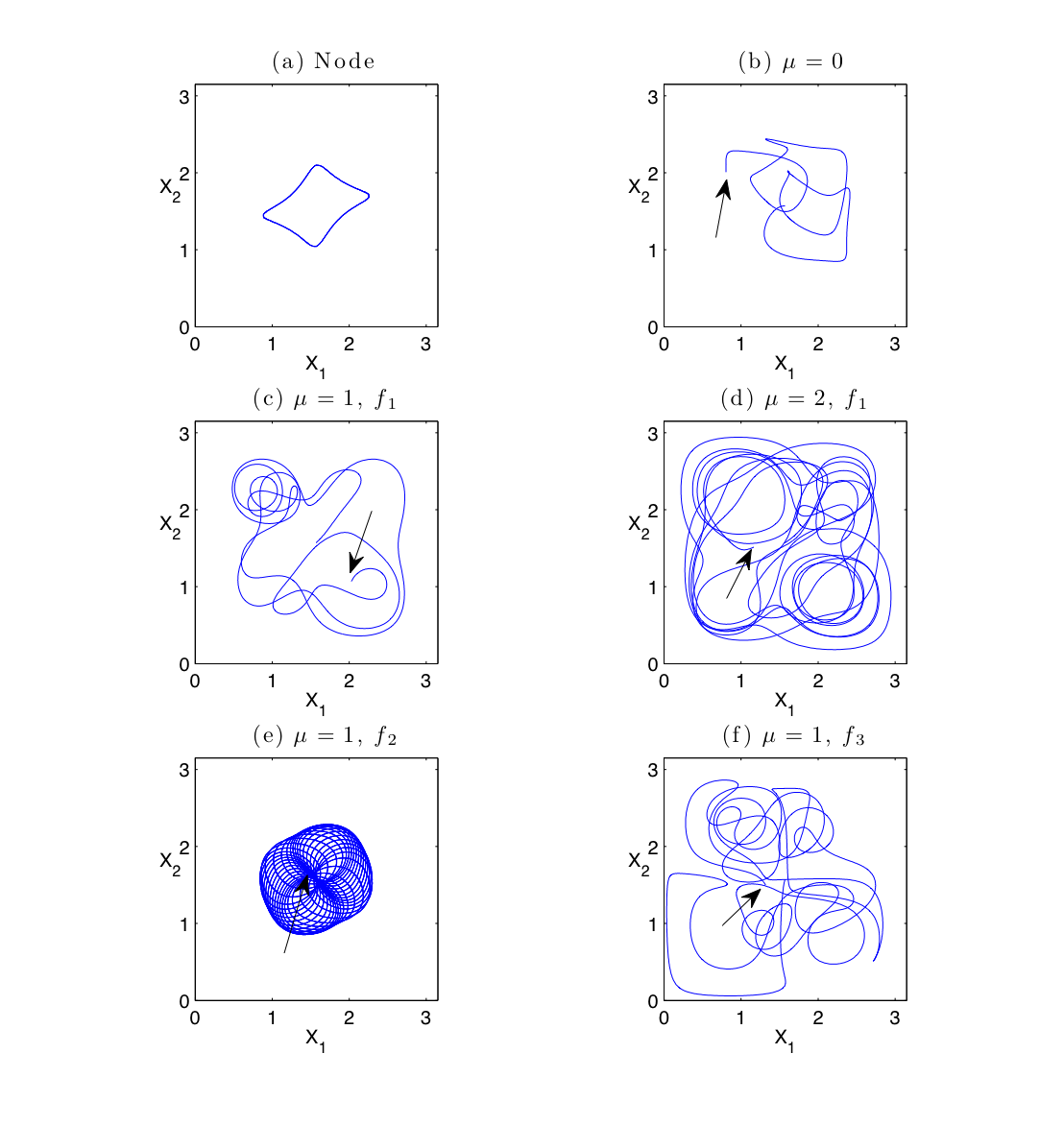}
\caption{Figure (a) displays the trajectory of the node of the wavefunction $\psi_1$. The other figures display particle trajectories for different values of $\mu$ and $f$, for the wavefunction $\psi_1$. In each case, the initial position is the center of the box. The final position is denoted by an arrow. All trajectories are for the time interval $[0,4\pi]$. Note that the trajectory in figure (e) does not display periodicity over this time interval.}
\label{trajectories1}
\end{figure}
\begin{figure}
\centering
\includegraphics[width=\textwidth]{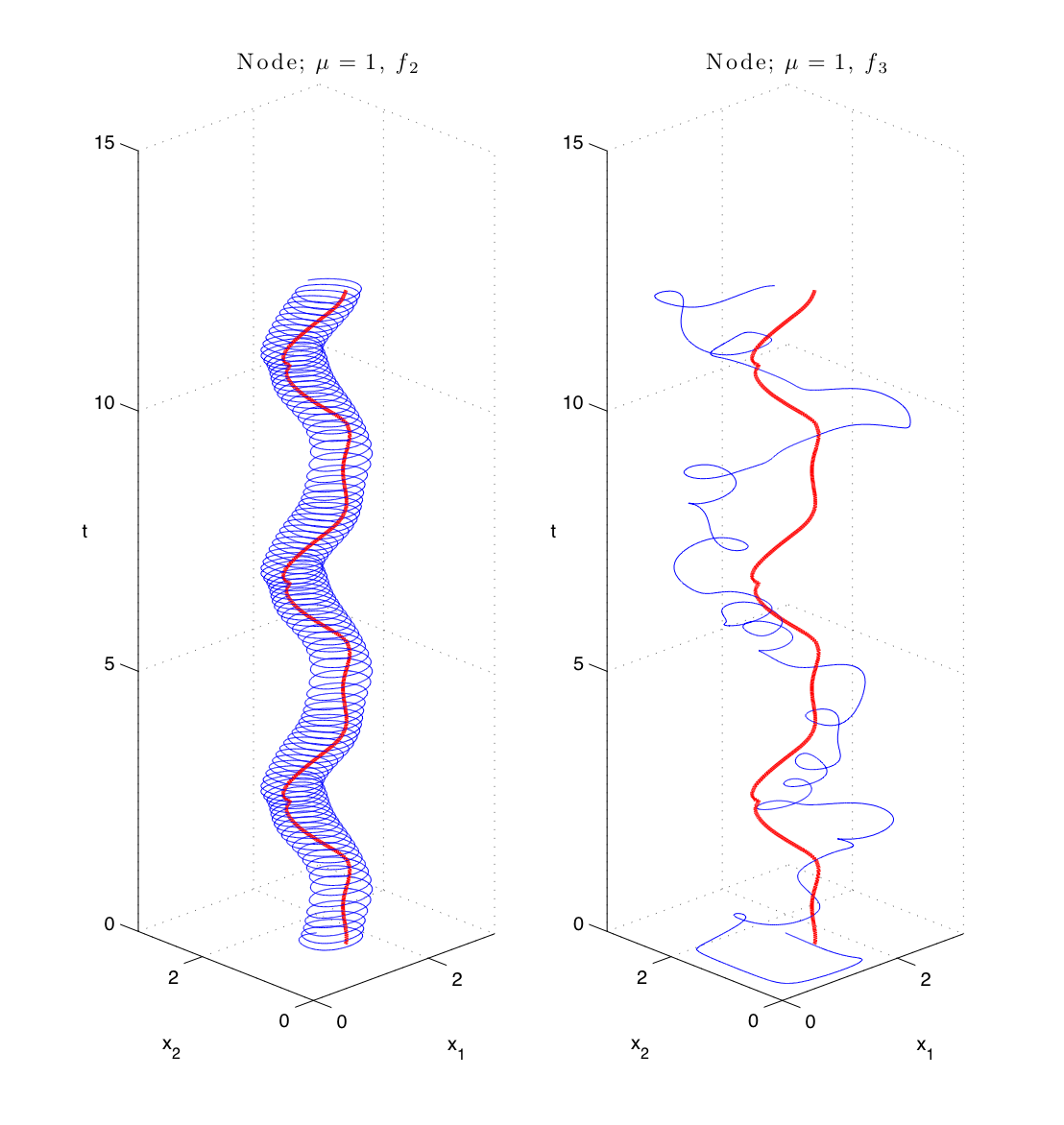}
\caption{Space-time trajectories of the node (bold lines) and particle trajectories (thin lines), for the wavefunction $\psi_1$, over the time interval $[0,4\pi]$. In each case, the initial position is the center of the box.}
\label{trajectories2}
\end{figure}

We consider different non-equilibrium distributions at time $t=0$. The first one, denoted by $\rho_0$, is chosen to be the square of the modulus of the ground state, that is,
\begin{equation}
\rho_0(x_1,x_2,0) = |\phi_{11}(x_1,x_2)|^2 = \left( \frac{2}{\pi} \right)^2 \sin^2x_1 \sin^2 x_2  \,.
\label{53}
\end{equation}
The other non-equilibrium distributions, denoted by $\rho_i$, $i=1,\dots,4$, are obtained from $\rho_0$ by translation and contraction, that is,
\begin{align}
\rho_1(x_1,x_2,0) &= 4\rho_0(2x_1,2x_2,0)\,, \quad & (x_1,x_2)\in[0,\pi/2]\times[0,\pi/2]\,, \nonumber\\
\rho_2(x_1,x_2,0) &= 4\rho_0(2(x_1-\pi/2),2x_2,0)\,, \quad & (x_1,x_2)\in[\pi/2,\pi]\times[0,\pi/2]\,,\nonumber\\
\rho_3(x_1,x_2,0) &= 4\rho_0(2x_1,2(x_2-\pi/2),0)\,, \quad & (x_1,x_2)\in[0,\pi/2]\times[\pi/2,\pi]\,,\nonumber\\
\rho_4(x_1,x_2,0) &= 4\rho_0(2(x_1-\pi/2),2(x_2-\pi/2),0)\,, \quad & (x_1,x_2)\in[\pi/2,\pi]\times[\pi/2,\pi]\,,
\label{54}
\end{align}
and are zero outside the domains specified. So these distributions have their support in the four different quadrants of the square box.

\subsection{Some details of the algorithm}
The algorithm to calculate the evolution of the non-equilibrium densities is similar to that of Valentini and Westman \cite{valentini05}. Just as in their work, the calculation of the trajectories is based on the Runge-Kutta-Fehlberg algorithm with Cash-Karp parameters \cite{press92} (but with a starting value of $\Delta$ equal to $10^{-5}$ and the maximal number of time steps equal to $10^5$).

In order to calculate the time evolved non-equilibrium density $\rho(x_1,x_2,t)$ from the initial density $\rho(x_1,x_2,0)$, a uniform lattice is placed over the box, whose lattice points have the coordinates $(k\pi/1024 - \pi/2048,l\pi/1024 - \pi/2048)$, where $k,l = 1,2,\ldots,1024$ (with hence a total of $1024 \times 1024$ lattice points). Each lattice point is then evolved backwards in time from time $t$ to time $t=0$ using the \dbb\ dynamics. Then, using the constancy of $f=\rho/|\psi|^2$ along a trajectory (see Section \ref{coarsegraining}), one has that the time evolved density at a lattice point with coordinates $(x_1,x_2)$ is given by $\rho(x_1,x_2,t) = |\psi(x_1,x_2,t)|^2 f(x_1(0),x_2(0),0)$, where $(x_1(0),x_2(0))$ are the coordinates of the backtracked position. This method of using the backtracked positions is advantageous compared to one where positions are evolved forward in time \cite{valentini05}. If the calculation of the backtracked position involves more than the $10^5$ time steps, it is halted and the lattice point is ignored when calculating the coarse-grained and smoothed density.

The coarse-graining of a density $\rho$ is done by averaging over square non-overlapping cells of side ${\pi}/{32}$ (with a total of $32 \times 32$ cells). Each such coarse-graining cell $C$ contains 1024 lattice points, of which a certain number $N_C$ can be backtracked. So the coarse-grained density ${\bar \rho}$ in a certain cell $C$ is given by $\sum_{C} \rho(x_1,x_2,t) / N_C$, where the sum ranges over coordinates $(x_1,x_2)$ of lattice points that are contained in the cell $C$ and that could be backtracked. 

Lattice points near nodes are generally harder to backtrack because the velocity field generally diverges there. This is in particular the case for the lattice points near the boundary of the box. Therefore in order to speed-up the calculation of the time evolved non-equilibrium densities, we have ignored those lattice points that lie within two coarse-graining cells near the boundary. As such there is an error in calculating the coarse-grained $H$-function, since the ignored regions near the boundary yield no contribution to the integral. Therefore our calculated coarse-grained $H$-function only measures the difference between the calculated distributions ${\bar \rho}$ and ${\overline{|\psi|^2}}$ over the region within the box that excludes the ignored coarse-graining cells.   

In the figures, smoothed densities ${\widetilde{\rho}}$ are displayed, instead of the coarse-grained densities ${\bar{\rho}}$. The reason for the smoothing is to average out fine-grained fluctuations. It is done by averaging over square cells of side ${\pi}/{16}$, ignoring again the cells near the boundary. The cells may be overlapping and can be mapped to one another by translations over an integer multiple of $\pi/108$ along the $x_1$ and $x_2$ directions. More explicitly, we consider the points with coordinates $(k\pi/128 + 3\pi/32,l\pi/128 + 3\pi/32)$, with $k,l = 0,1,2,\ldots,104$, and calculate ${\widetilde{\rho}}$ at those points by averaging the density $\rho$ over cells of side ${\pi}/{16}$, ignoring again the lattice points that could not be backtracked (so that each average is done over at most 4096 lattice points). In generating the plots with Matlab an interpolation of those densities values is then introduced.

\subsection{Results of the numerical simulations}
Figures \ref{fig1}-\ref{fig4} display the time evolution of the distributions $|\psi_1|^2$ and of the different non-equilibrium distributions $\rho_i$, $i=0,\dots,4$ for $\psi_1$ (given by \eqref{51}), and this for different guidance equations \eqref{52} that are distinguished by different values of $\mu$ and $f$. Figure \ref{fig5} displays the distributions for the nodeless wavefunction $\psi_2$ (given in \eqref{51.1}).

\begin{figure}
\centering
\includegraphics[width=\textwidth]{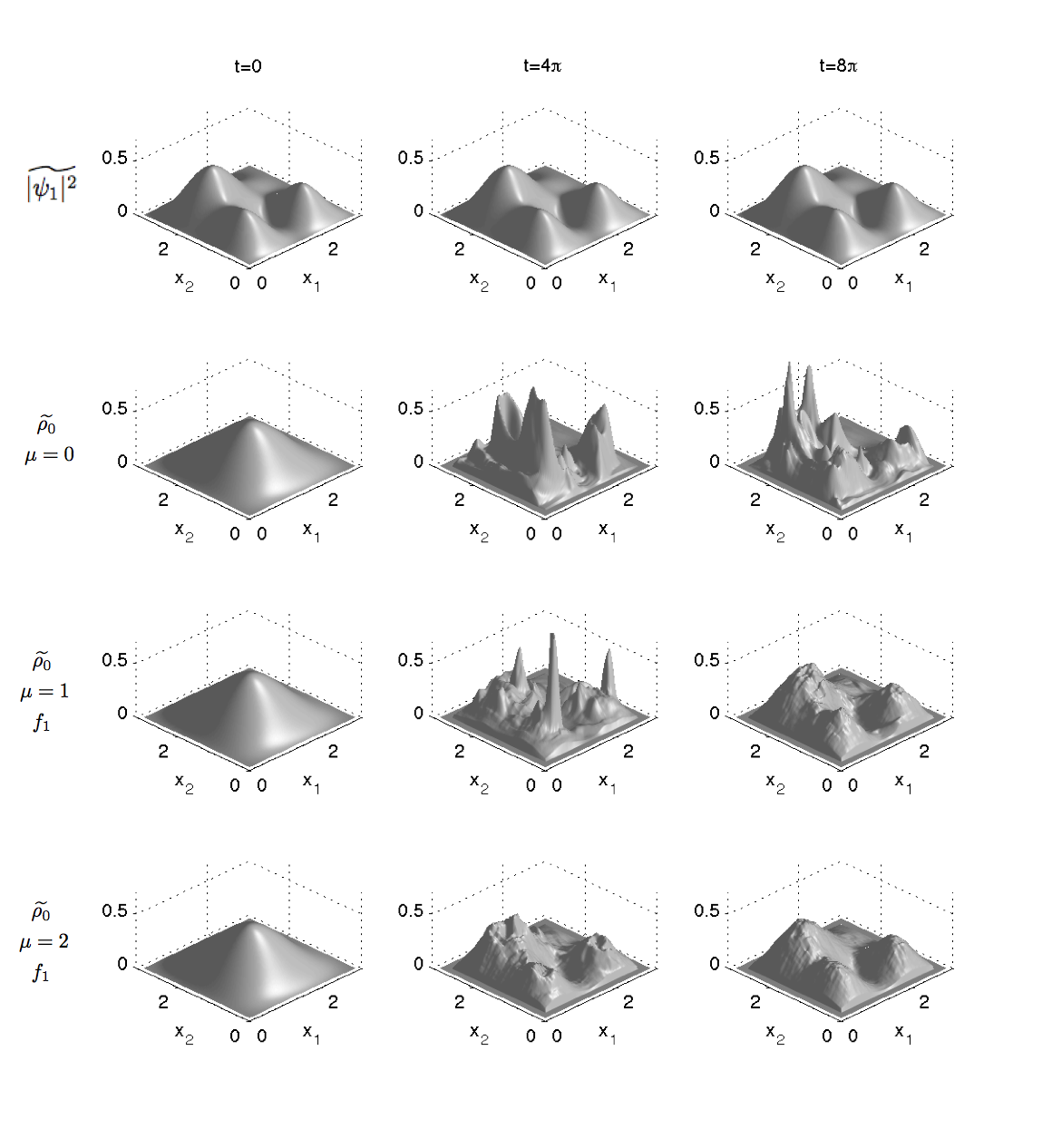}
\caption{Smoothed densities ${\widetilde{|\psi_1|^2}}$ and ${\widetilde \rho}_0$ at times $t=0$, $t=4\pi$ and $t=8\pi$, for guidance equations with different values of $\mu$ and $f=f_1$, for the wavefunction $\psi_1$. Notice that the relaxation to equilibrium is better for higher value of $\mu$.}
\label{fig1}
\end{figure}

\begin{figure}
\centering
\includegraphics[width=\textwidth]{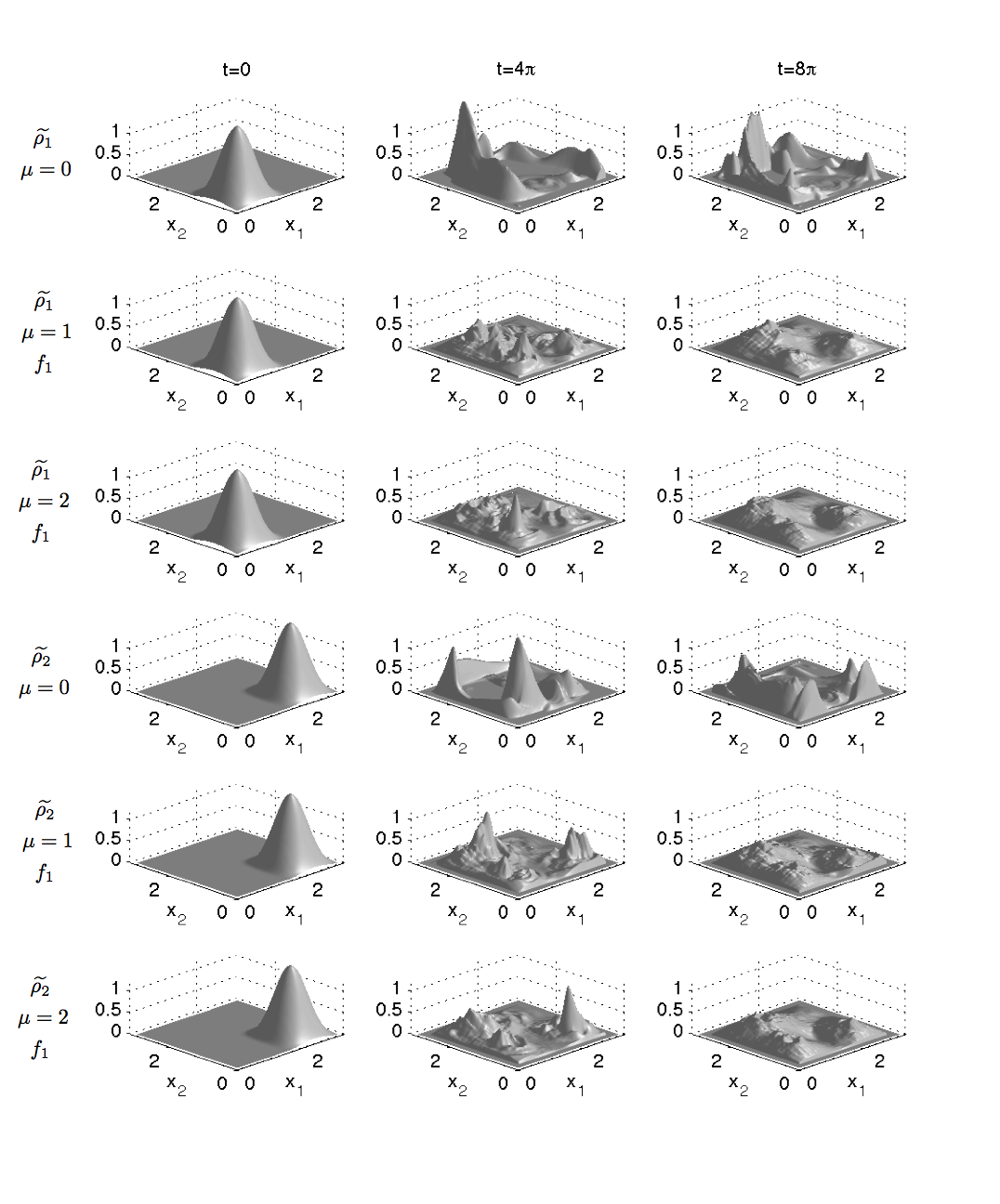}
\caption{Smoothed densities ${\widetilde \rho}_1$ and ${\widetilde \rho}_2$ at times $t=0$, $t=4\pi$ and $t=8\pi$, for guidance equations with different values of $\mu$ and $f=f_1$, for the wavefunction $\psi_1$.}
\label{fig2}
\end{figure}

\begin{figure}
\centering
\includegraphics[width=\textwidth]{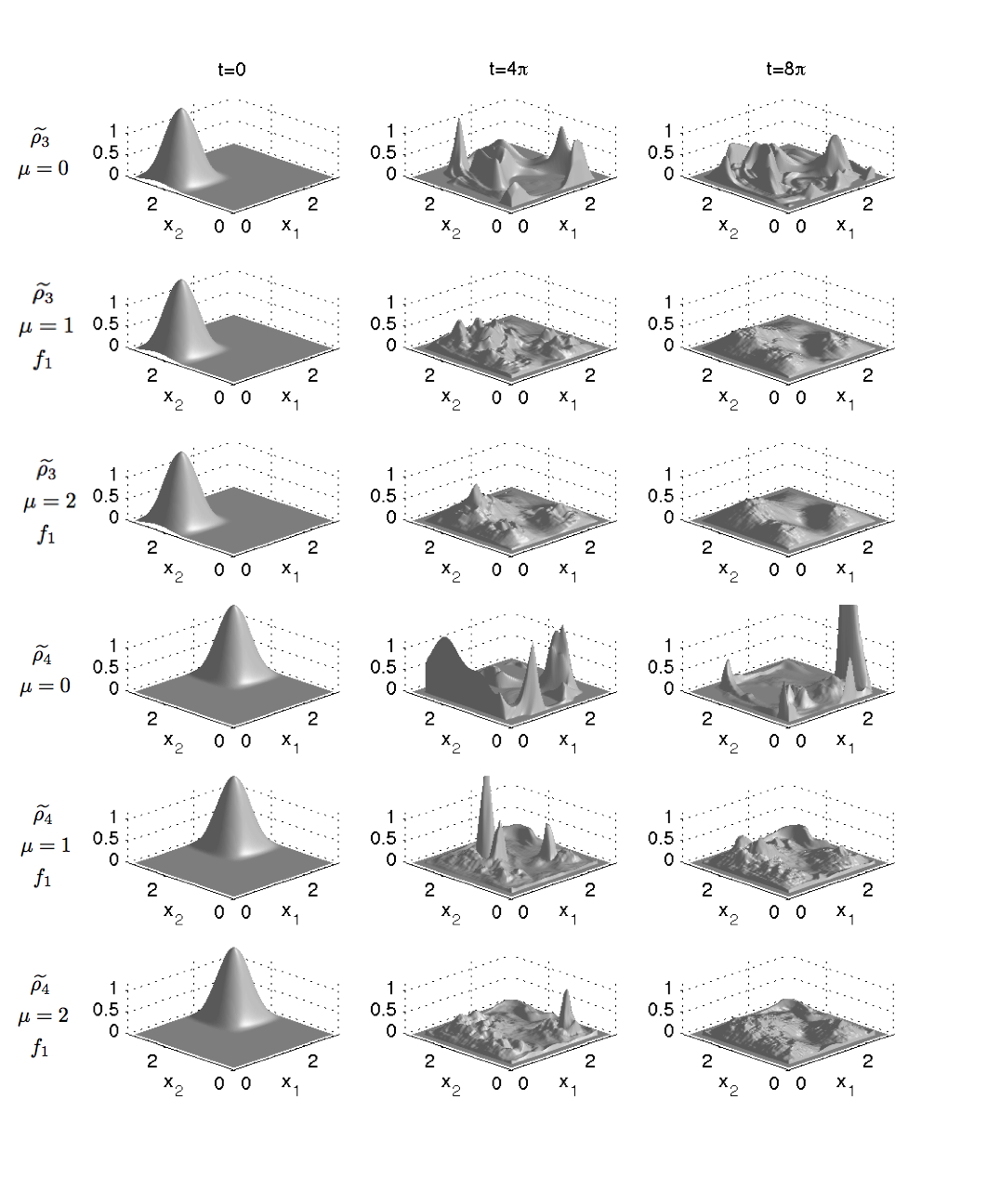}
\caption{Smoothed densities ${\widetilde \rho}_3$ and ${\widetilde \rho}_4$ at times $t=0$, $t=4\pi$ and $t=8\pi$, for guidance equations with different values of $\mu$ and $f=f_1$, for the wavefunction $\psi_1$. (In the case of ${\widetilde \rho}_4$, $\mu=0$ and $t=8\pi$, one peak in the distribution is not fully plotted.)}
\label{fig3}
\end{figure}

\begin{figure}
\centering
\includegraphics[width=\textwidth]{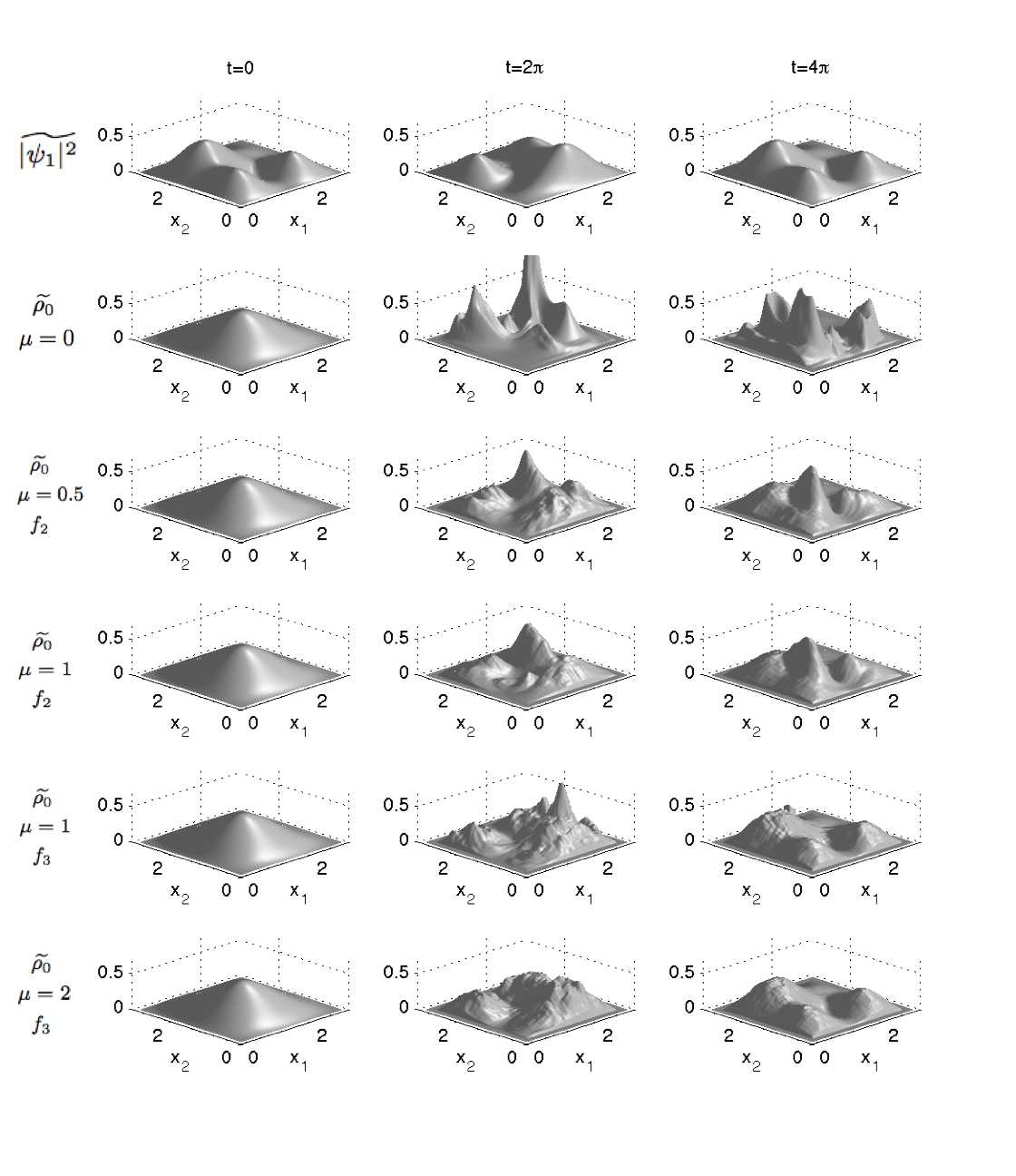}
\caption{Smoothed densities ${\widetilde{|\psi_1|^2}}$ and ${\widetilde \rho}_0$ at times $t=0$, $t=2\pi$ and $t=4\pi$, for guidance equations with different values of $\mu$ and $f$, for the wavefunction $\psi_1$. (In the case of ${\widetilde \rho}_0$, $\mu=0$ and $t=2\pi$, one peak in the distribution is not fully plotted.)}
\label{fig4}
\end{figure}

\begin{figure}
\centering
\includegraphics[width=\textwidth]{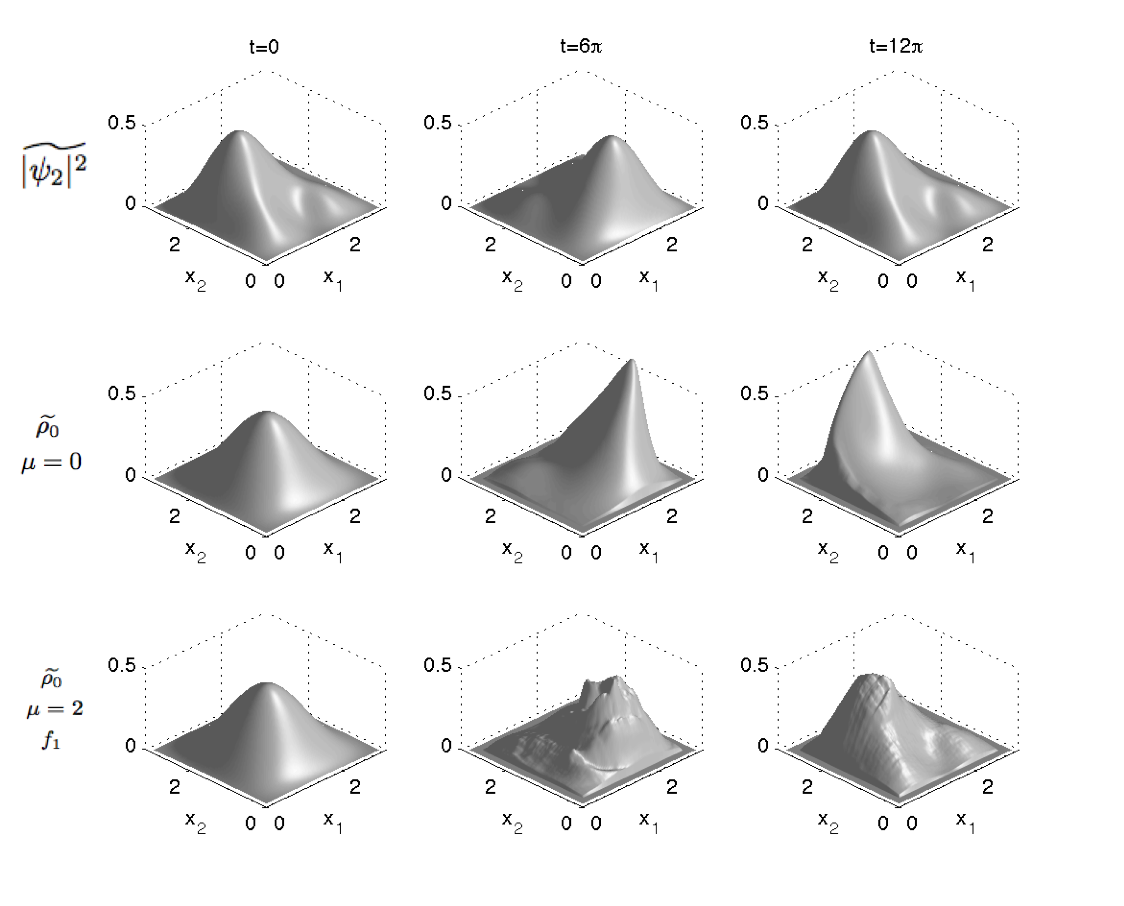}
\caption{Smoothed densities ${\widetilde{|\psi_2|^2}}$ and ${\widetilde \rho}_0$ at times $t=0$, $t=6\pi$ and $t=12\pi$, for the ordinary \dbb\ guidance equation and for the one with $\mu=2$ and $f=f_1$, for the wavefunction $\psi_2$.}
\label{fig5}
\end{figure}

The percentages of lattice points that could be backtracked in each case are given in the tables \ref{goodtrajectoriesnode} and \ref{goodtrajectoriesnodeless}. In most cases the percentages over coarse-graining cells did not differ all too much from the percentages over the box. The coarse-graining cells with the worst percentages were generally located near the boundary of the box.

\begin{table}
\begin{center}
\begin{tabular}{l | r @{.} l   r @{.} l  }
                          & \multicolumn{2}{c}{$t=4\pi$} & \multicolumn{2}{c}{$t=8\pi$} \\
\hline 
$\mu=0$                   &   99&89                      &  99&62 \\
$\mu=1$, $f_1$            &  99&98                        &  99&52 \\
$\mu=2$, $f_1$            &   99&52                      &  94&44 \\
\end{tabular}
\begin{tabular}{c}
\qquad \qquad
\end{tabular}
\begin{tabular}{l | r @{.} l   r @{.} l  }
                          & \multicolumn{2}{c}{$t=2\pi$} & \multicolumn{2}{c}{$t=4\pi$}\\
\hline
$\mu=0$                   &   99&95                       &  99&89 \\
$\mu=\frac{1}{2}$, $f_2$  &  100&                        &  99&62 \\
$\mu=1$, $f_2$            &   99&99                      &  91&71 \\
$\mu=1$, $f_3$            &  100&                         &  99&82 \\
$\mu=2$, $f_3$            &   99&99                       &  97&16\\
\end{tabular}
\caption{Percentage of backtracked lattice points, for the wavefunction $\psi_1$, and for the guidance equations with different values of $\mu$ and $f$ (up to two decimal places).}
\label{goodtrajectoriesnode}
\end{center}

\begin{center}
\begin{tabular}{l | r @{.} l   r @{.} l  }
                          & \multicolumn{2}{c}{$t=6\pi$} & \multicolumn{2}{c}{$t=12\pi$}\\
 \hline
$\mu=0$                   &  100&                         & 100& \\
$ \mu=2$, $f_1$           &   99&45                       &  89&21 \\
\end{tabular}
\caption{Percentage of backtracked lattice points, for the nodeless wavefunction $\psi_2$, for the ordinary \dbb\ guidance equation and for the one with $\mu=2$ and $f=f_1$ (up to two decimal places).}
\label{goodtrajectoriesnodeless}
\end{center}
\end{table}

The calculated values of the coarse-grained $H$-function are given in tables \ref{hfunctionnode} and \ref{hfunctionnodeless}. Note that some values of ${\bar H}$ are negative, which is theoretically impossible. In general there will be some deviation from the actual values, because of several reasons. First, these densities are calculated using only a lattice sampling, second there are the numerical errors that come up in calculating the time evolution of the distributions, and third, by ignoring the contributions of the regions near the boundary in the integral that defines ${\bar H}$, we have effectively put ${\bar \rho}$ zero in those regions. In \cite{valentini05} Valentini and Westman estimated the error in ${\bar H}$ to be about $2\%$. Assuming a similar error in our case, we have presented the values of ${\bar H}$ rounded to the nearest integer.

\begin{table}
\begin{center}
\begin{tabular}{l | r    r    r }
                          & \multicolumn{1}{c}{$t=0$} & \multicolumn{1}{c}{$t=4\pi$}& \multicolumn{1}{c}{$t=8\pi$}\\
\hline
$\bar\rho_0$, $\mu=0 $        &    54         &     31      &   22 \\
$\bar\rho_0$, $\mu=1$, $f_1$  &    54         &     10      &    1 \\
$\bar\rho_0$, $\mu=2$, $f_1$  &    54        &      1      &   -1 \\
$\bar\rho_1$, $\mu=0 $        &   168         &    126      &   90 \\
$\bar\rho_1$, $\mu=1$, $f_1$  &   168        &     16      &    2 \\
$\bar\rho_1$, $\mu=2$, $f_1$  &   168         &     14      &   0 \\
$\bar\rho_2$, $\mu=0 $        &   169         &    129     &   91 \\
$\bar\rho_2$, $\mu=1$, $f_1$  &   169         &     26      &    5 \\
$\bar\rho_2$, $\mu=2$, $f_1$  &   169         &     17      &    1 \\
$\bar\rho_3$, $\mu=0 $        &   130         &     94      &   61\\
$\bar\rho_3$, $\mu=1$, $f_1$  &   130         &     12       &    1 \\
$\bar\rho_3$, $\mu=2$, $f_1$  &   130         &      5      &   0 \\
$\bar\rho_4$, $\mu=0$         &   300         &    207      &  158 \\
$\bar\rho_4$, $\mu=1$, $f_1$  &   300         &     54       &   19 \\
$\bar\rho_4$, $\mu=2$, $f_1$  &   300         &     21      &    7 \\
\end{tabular}
\begin{tabular}{l |  r    r    r  }
                          & $t=0$ & $t=2\pi$      & $t=4\pi$\\
\hline
$\bar\rho_0$, $\mu=0$                   & 54        &        39    &  31 \\
$\bar\rho_0$, $\mu=\frac{1}{2}$, $f_2$  & 54        &        23   &  21 \\
$\bar\rho_0$, $\mu=1$, $f_2$            & 54        &        22    &  18 \\
$\bar\rho_0$, $\mu=1$, $f_3$            & 54        &         5   &   1 \\
$\bar\rho_0$, $\mu=2$, $f_3$            & 54        &         2   &  -1\\
\end{tabular}
\caption{Values of ${\bar H}$ (rounded to the nearest integer) for the densities plotted in figures \ref{fig1}-\ref{fig4}. Notice the decrease of ${\bar H}$ with increase of $\mu$.} 
\label{hfunctionnode}
\end{center}

\begin{center}
\begin{tabular}{l | r    r    r }
                          & $t=0$ & $t=6\pi$ & $t=12\pi$\\
\hline
$\bar\rho_0~ \mu=0$         &     17     &   17   &    16 \\
$\bar\rho_0~ \mu=2$, $f_1$  &     17     &    1   &    -1 \\
\end{tabular}
\caption{Values of ${\bar H}$ (rounded to the nearest integer) for the densities plotted in figure \ref{fig5}.} 
\label{hfunctionnodeless}
\end{center}
\end{table}

From the pictures, as well as from the values of the coarse-grained $H$-function, it is clear that the non-equilibrium distributions evolve closer to equilibrium over time. Furthermore, the relaxation time decreases when the value of $\mu$ increases. As mentioned before, this might be explained by the fact that the extra term in the velocity adds to the overall irregularity of the motion and to the vorticity in particular. 

Note that in the case of the ordinary \dbb\ velocity field, the relaxation is not as good as in the simulation of Valentini and Westman \cite{valentini05}, where a superposition of the first 16 modes was considered instead of the first 4. This is probably related to the fact that more chaotic behaviour is expected when the number of nodes increases \cite{frisk97}. The wavefunction of Valentini and Westman namely has much more than one node. 

In the case of the guidance equation with $f=f_2$ there is not a very good relaxation neither for $\mu=0.5$, nor for $\mu=1$, compared to the other alternative guidance equation equations (though it is still better than for the ordinary \dbb\ theory). We have tried to perform the simulations also for $\mu=1.5$, but found that only $76.78\%$ of the lattice points could be backtracked. Since this percentage is rather low, we did not include these results in the paper. Figures \ref{trajectories1} and \ref{trajectories2} display some trajectories for different possible guidance equations. The trajectory for $f=f_2$ and $\mu=1$ circles around the node from time $t=0$ to $t=4\pi$, unlike the trajectories corresponding to the other guidance equations, which tend to cover a bigger area of the box in that time. This might be a possible reason why the relaxation is not so good in this case. If the dynamics is such that initial positions that start near a node remain close to it for long periods of time and merely circle around it, one might expect that there is insufficient mixing-like behaviour for relaxation to occur efficiently. 

In the case of the wavefunction without node, there is some convergence towards quantum equilibrium for the ordinary \dbb\ guidance equation. However, there is definitely not a good relaxation. It is unclear how much better the convergence would be for larger times. In the case of the guidance equation with $\mu=2$ and $f=f_1$ there is an approximate relaxation to equilibrium. Presumably the more efficient relaxation in the latter case is related to possible chaotic behaviour induced by the vorticity. In the ordinary \dbb\ theory there is probably no chaotic motion for this wavefunction, due to the absence of nodes.

\section{Conclusion}
We have considered non-equilibrium distributions and their relaxation properties for a class of \dbb-type theories, and found that the relaxation time depends substantially on the form of the guidance equation. In particular, in the case of the examples of such theories considered here, there was a faster relaxation compared to the ordinary \dbb\ theory. 

Only non-relativistic quantum theory was considered, but similar results are expected to hold for relativistic quantum theory or quantum field theory. As such, these results might have potential implications for empirical predictions concerning non-equilibrium distributions, like for example those of Valentini \cite{valentini08a,valentini08b} that were mentioned in the introduction. But, with this being said, Valentini's predictions involve a field ontology for a scalar field, with a velocity field that is most natural. In support of his choice, Valentini actually proves uniqueness of this velocity field for the vacuum state (the state of interest for his analysis) under certain natural assumptions \cite{valentini08b}.

Actually, when it comes to quantum field theory, there is, apart from the ambiguities in the velocity field, also a potential ambiguity in the choice of ontology. Some \dbb-type models for example suggest a field ontology, while others suggest a particle ontology (see for example \cite{durr04,colin07} for possible particle ontologies and \cite{struyve07a} for a review on possible field ontologies). Such ambiguities may potentially be relevant too for possible empirical predictions concerning non-equilibrium.  

Note that similar ambiguities arise in for example the context of Nelson's stochastic mechanics \cite{davidson79,bacciagaluppi99}. Nelson's mechanics \cite{nelson66,goldstein87} can be regarded as a modification of the \dbb\ theory in which the deterministic \dbb\ dynamics is replaced by a diffusion process (though Nelson himself wanted to derive the wavefunction and Schr\"odinger's dynamics from the stochastic process instead of assuming them independently). This diffusion process is expected to yield convergence to equilibrium for most distributions (see \cite{cufaropetroni98} for some rigorous results). As with the \dbb\ theory there are many ways in which a Nelsonian type of dynamics can be introduced that preserves the equilibrium distribution $|\psi|^2$. In particular, there are Nelsonian theories with arbitrary values of the diffusion constant \cite{davidson79}. In this case, relaxation time is expected to decrease with increasing diffusion constant.

\section{Acknowledgments}
S.C.\ acknowledges support from a PIAF fellowship (Perimeter Institute Australia Foundations) and is grateful to Josh Combes, Indy Siva and Mike Towler for their help with some aspects regarding the numerical simulations.  W.S.\ acknowledges support of the FWO-Flanders and is grateful to Lukas Maes and Ruben Stukken for their help with the initial exploration of this project, and to Christian Maes for useful discussions. We further thank Antony Valentini, Hans Westman and Howard Wiseman for valuable discussions or comments. It is also a pleasure to thank Perimeter Institute for hospitality enjoyed during a stay where part of this work was completed (W.S.\ thanks Lucien Hardy for the invitation).


\begin{thebibliography}{10}
\bibitem{bohm93}
{D.\ Bohm and B.J.\ Hiley, {\em The Undivided Universe}, Routledge, New York
  (1993).}

\bibitem{holland93b}
{P.R.\ Holland, {\em The Quantum Theory of Motion}, Cambridge University Press,
  Cambridge (1993).}

\bibitem{durr09}
{D.\ D\"urr and S.\ Teufel, {\em Bohmian Mechanics}, Springer-Verlag, Berlin
  (2009).}

\bibitem{durr92}
{D.\ D\"urr, S.\ Goldstein and N.\ Zangh\`\i, {\em J.\ Stat.\ Phys.}\ {\bf 67},
  843 (1992) and quant-ph/0308039.}

\bibitem{bohm53b}
{D.\ Bohm, {\em Phys.\ Rev.}\ {\bf 89}, 458 (1953).}

\bibitem{valentini91a}
{A.\ Valentini, {\em Phys.\ Lett.\ A} {\bf 156}, 5 (1991).}

\bibitem{valentini05}
{A.\ Valentini and H.\ Westman, {\em Proc.\ R.\ Soc.\ A} {\bf 461}, 253 (2005)
  and quant-ph/0403034.}

\bibitem{valentini92}
{A.\ Valentini, {\em On the Pilot-Wave Theory of Classical, Quantum and
  Subquantum Physics}, PhD.\ Thesis, International School for Advanced Studies,
  Trieste (1992), online
  \url{http://www.sissa.it/ap/PhD/Theses/valentini.pdf}.}

\bibitem{valentini04a}
{A.\ Valentini, {\em Phys.\ Lett.\ A} {\bf 332}, 187 (2004) and
  quant-ph/0309107.}

\bibitem{valentini07}
{A.\ Valentini, {\em J.\ Phys.\ A} {\bf 40}, 3285 (2007) and hep-th/0610032.}

\bibitem{valentini08a}
{A.\ Valentini, 0804.4656 [hep-th].}

\bibitem{valentini08b}
{A.\ Valentini, 0805.0163 [hep-th].}

\bibitem{deotto98}
{E.\ Deotto and G.C.\ Ghirardi, {\em Found.\ Phys.}\ {\bf 28}, 1 (1998) and
  quant-ph/9704021.}

\bibitem{frisk97}
{H.\ Frisk, {\em Phys.\ Lett.\ A} {\bf 227}, 139 (1997).}

\bibitem{wu99}
{H.\ Wu and D.W.L. Sprung, {\em Phys.\ Lett.\ A} {\bf 261}, 150 (1999).}

\bibitem{wisniacki05}
{D.A.\ Wisniacki and E.R.\ Pujals, {\em Europhys. Lett.}\ {\bf 71}, 159 (2005)
  and quant-ph/0502108.}

\bibitem{wisniacki06}
{D.A.\ Wisniacki, E.R.\ Pujals and F.\ Borondo, {\em Europhys. Lett.}\ {\bf
  73}, 671 (2006) and nlin/0507015.}

\bibitem{efthymiopoulos07}
{C.\ Efthymiopoulos, C.\ Kalapotharakos and G.\ Contopoulos, {\em J.\ Phys.\ A}
  {\bf 40}, 12945 (2007) and 0709.2038 [quant-ph].}

\bibitem{efthymiopoulos09}
{C.\ Efthymiopoulos, C.\ Kalapotharakos and G.\ Contopoulos, {\em Phys.\ Rev.\
  E} {\bf 79}, 036203 (2009) and 0903.2655 [quant-ph].}

\bibitem{wyatt05}
{R.E.\ Wyatt, {\em Quantum Dynamics with Trajectories}, Springer, New York
  (2005).}

\bibitem{goldstein07}
{S.\ Goldstein and W.\ Struyve, {\em J.\ Stat.\ Phys.}\ {\bf 128}, 1197 (2007)
  and 0704.3070 [quant-ph].}

\bibitem{bohm52b}
{D.\ Bohm, {\em Phys.\ Rev.}\ {\bf 85}, 180 (1952).}

\bibitem{durr033}
{D.\ D\"urr, S.\ Goldstein and N.\ Zangh\`\i, {\em J.\ Stat.\
  Phys.}\ {\bf 116}, 959 (2004) and quant-ph/0308038.}

\bibitem{holland99}
{P.R.\ Holland, {\em Phys.\ Rev.\ A} {\bf 60}, 4326 (1999).}

\bibitem{holland03b}
{P.R.\ Holland, {\em Ann.\ Phys.\ (Leipzig)} {\bf 12}, 446 (2003) and
  quant-ph/0305175.}

\bibitem{struyve03b}
{W.\ Struyve, W.\ De Baere, J.\ De Neve and S.\ De Weirdt, {\em Phys.\ Lett.\
  A} {\bf 322}, 84 (2004) and quant-ph/0311098.}

\bibitem{wiseman07}
{H.M.\ Wiseman, {\em New J.\ Phys.}\ {\bf 9}, 165 (2007) and 0706.2522
  [quant-ph].}

\bibitem{holland03a}
{P.R.\ Holland and C.\ Philippidis, {\em Phys.\ Rev.\ A} {\bf 67}, 062105
  (2003) and quant-ph/0302076.}

\bibitem{colijn02}
{C.\ Colijn and E.R.\ Vrscay, {\em Phys.\ Lett.\ A} {\bf 300}, 334 (2002) and
  quant-ph/0308105.}

\bibitem{colijn03a}
{C.\ Colijn and E.R.\ Vrscay, {\em J.\ Phys.\ A} {\bf 36}, 4689 (2003) and
  quant-ph/0304203.}

\bibitem{colijn03b}
{C.\ Colijn and E.R.\ Vrscay, {\em Found.\ Phys.\ Lett.}\ {\bf 16}, 303 (2003)
  and quant-ph/0304198.}

\bibitem{struyve05}
{W.\ Struyve, {\em The de Broglie-Bohm pilot-wave interpretation of quantum
  theory}, PhD.\ Thesis, Ghent University, Ghent (2004) and quant-ph/0506243.}

\bibitem{holland98}
{P.R.\ Holland, {\em Found.\ Phys.}\ {\bf 28}, 881 (1998).}

\bibitem{goldstein042}
{S.\ Goldstein, J.\ Taylor, R.\ Tumulka and N.\ Zangh\`\i, {\em J.\ Phys.\ A}
  {\bf 38}, 1567 (2005) and quant-ph/0405039.}

\bibitem{struyve09b}
{W.\ Struyve, 0907.5373 [quant-ph].}

\bibitem{berndl93}
{K.\ Berndl, D.\ D\"urr, S.\ Goldstein and N.\ Zangh\`\i, {\em Int.\ J.\
  Theor.\ Phys.}\ {\bf 32}, 2245 (1993).}

\bibitem{teufel04}
{S.\ Teufel and R.\ Tumulka, {\em Commun.\ Math.\ Phys}.\ 258 (2005) 349-365
  and math-ph/0406030 (2004).}

\bibitem{takabayasi83}
{T.\ Takabayasi, {\em Prog.\ Theor.\ Phys.}\ {\bf 69}, 1323 (1983).}

\bibitem{dirac31}
{P.A.M.\ Dirac, {\em Proc.\ R.\ Soc.\ A} {\bf 133}, 60 (1931).}

\bibitem{hirschfelder74}
{J.O.\ Hirschfelder, C.J.\ Goebel and L.W.\ Bruch, {\em J.\ Chem.\ Phys.}\ {\bf
  61}, 5456 (1974).}

\bibitem{colijn04}
{C.\ Colijn and E.R.\ Vrscay, {\em Phys.\ Lett.\ A} {\bf 327}, 113 (2004).}

\bibitem{walters82}
{P.\ Walters, {\em An Introduction to Ergodic Theory}, {\em Graduate text in
  mathematics} {\bf 79}, Springer-Verlag, New-York (1982).}

\bibitem{goldstein99}
{S.\ Goldstein, {\em Phys.\ Rev.\ E} {\bf 60}, 7578 (1999) and
  quant-ph/9901005.}

\bibitem{press92}
{W.H.\ Press, S.A.\ Teukolsky, W.T.\ Vetterling and B.P.\ Flannery, {\em
  Numerical recipes in FORTRAN}, Cambridge University press, Cambridge (1992).}

\bibitem{durr04}
{D.\ D\"urr, S.\ Goldstein, R.\ Tumulka and N.\ Zangh\`\i, {\em J.\ Phys.\ A}
  {\bf 38}, R1 (2005) and quant-ph/0407116.}

\bibitem{colin07}
{S.\ Colin and W.\ Struyve, {\em J.\ Phys.\ A} {\bf 40}, 7309 (2007) and
  quant-ph/0701085.}

\bibitem{struyve07a}
{W.\ Struyve, 0707.3685 [quant-ph].}

\bibitem{davidson79}
{M.\ Davidson, {\em Lett.\ Math.\ Phys.}\ {\bf 3}, 271 (1979).}

\bibitem{bacciagaluppi99}
{G.\ Bacciagaluppi, {\em Found.\ Phys.\ Lett} {\bf 12}, 1 (1999) and
  quant-ph/9811040.}

\bibitem{nelson66}
{E.\ Nelson, {\em Phys.\ Rev.}\ {\bf 150}, 1079 (1966).}

\bibitem{goldstein87}
{S.\ Goldstein, {\em J.\ Stat.\ Phys.}\ {\bf 47}, 645 (1987).}

\bibitem{cufaropetroni98}
{N.\ Cufaro Petroni, S.\ De Martino and S.\ De Siena, {\em Phys.\ Lett.\ A}
  {\bf 245}, 1 (1998).}

\end{thebibliography}
\end{document}